\documentclass[11pt]{article}
\usepackage{helvet}
\usepackage{url}
\usepackage{a4,graphicx}
\usepackage{natbib}
\usepackage[latin1]{inputenc}
\usepackage{amssymb}
\usepackage{amsmath}
\usepackage{paralist}
\usepackage{bibspacing}
\usepackage{chapterbib}
\usepackage{mathrsfs}
\usepackage{wasysym}
\usepackage{array}
\usepackage{makeidx}
\usepackage{floatflt}
\usepackage{dcolumn} 
\usepackage{multicol}
\usepackage{ifthen}
\usepackage{eurosym}
\usepackage{sidecap}
\usepackage{hyperref}
\renewcommand{\refname}{3~~~Project- and subject-related list of publications} 

\begin{document}

\begin{center}

{\Large \bf ESO Expanding Horizons White Paper}
\vspace{0.5cm}

{\Huge \bf Experiments in binary evolution}\\
\vspace{1.0cm}

{\Large Stephan Geier$^1$, Thomas Kupfer$^2$, Pierre Maxted$^3$, Veronika Schaffenroth$^4$}\\
\end{center}
\vspace{0.5cm}
{\small $^1$Institut f\"ur Physik und Astronomie, Universit\"at Potsdam, Haus 28, Karl-Liebknecht-Str. 24/25, D-14476 Potsdam-Golm, Germany}\\
{\small $^2$Hamburger Sternwarte, University of Hamburg, Gojenbergsweg 112, 21029 Hamburg, Germany}\\
{\small $^3$Astrophysics Group, Keele University, Staffordshire ST5 5BG, UK}\\
{\small $^4$Th\"uringer Landessternwarte Tautenburg, Sternwarte 5, 07778 Tautenburg, Germany}\\

The majority of stars more massive than the Sun is found in binary or multiple star systems and many of them will interact during their evolution. Specific interactions, where progenitors and post-mass transfer (MT) systems are clearly linked, can provide yet missing observational constraints. Volume-complete samples of progenitor and post-MT systems are well suited to study those processes. To compile them, we need to determine the parameters of thousands of binary systems with periods spanning several orders of magnitude. The bottleneck are the orbital parameters, because accurate determinations require a good coverage of the orbital phases. The next generation of time-resolved spectroscopic surveys should be optimized to follow-up and solve whole populations of binary systems in an efficient way. To achieve this, a scheduler predicting the best times of the next observation for any given target in real time should be combined with a flexibly schedulable multi-object spectrograph or ideally a network of independent telescopes.

\newpage

\section{Studying binary evolution with volume-complete samples}

The majority of stars more massive than the Sun is found in binary or multiple star systems and also less massive stars have significant multiplicity fractions ($\sim25\%$). A significant fraction of the systems is close enough that binary interactions will happen at some point of their evolution (Offner et al. 2023, ASP Conf. Ser., 534, 275). These interactions are in most cases initiated, when a star expands as it evolves. As soon as the radius of this star $R_{\rm donor}$ exceeds it's Roche radius $R_{\rm Roche}$, mass transfer to the companion starts (Roche Lobe Overflow, RLOF). What happens next, is a crucial topic in theoretical stellar astrophysics (Marchant \& Bodensteiner 2024, ARA\&A, 62, 21; Mathieu \& Pols 2025, ARA\&A, 63, 467). The main paradigm is that the mode of mass transfer is dictated by the mass ratio between the donor and the accretor star $q=M_{\rm donor}/M_{\rm accretor}$ as well as the internal structure of the donor (e.g. Temmink et al. 2023, A\&A, 669, A45). For small values of $q$, mass transfer is predicted to be stable. However, it is unclear, how much mass and angular momentum are transferred to the companion and how much are lost from the system. These processes determine how the masses and orbital parameters of the binary change during the interaction.

For higher $q$, the mass transfer is predicted to be unstable leading to a common envelope (CE) phase (Ivanova et al. 2020, AAS-IOP Astronomy Book Series). Possible outcomes are either an ejection of the CE or the merger of the two stars. Since the CE phase is accompanied by a spiral-in of the components, the resulting binaries are expected to have shorter periods than the initial binaries. Unfortunately, current models do not allow to make any more specific predictions about the parameters of post-CE systems based on the underlying physics alone. Simplified models using empirically calibrated parameters have to be used (Han et al. 2020, RAA, 20, 161). Critical mass ratios $q_{\rm crit}$ are predicted to mark the border between stable and unstable mass transfer. {\it The stability criteria are derived based on theoretical models, but there are very few and sometimes inconsistent constraints from observations.}

What we need to constrain the models are observations of the progenitor system of a specific interaction and the binary or single star, which is the result of the interaction. In this way we could directly determine the changes in the relevant parameters. To get an idea about those evolutionary connections, the space densities $\rho$ of stars are used as observational probes. To determine $\rho$ one needs to count all the stars with certain properties within a certain volume. The Gaia mission (Gaia Collaboration 2016, A\&A, 595, A1) is a major game changer in this respect since it provides accurate parallax-based distances to about $10^{7}$ stars within a few kpc around the Sun. The main advantage of volume limits is that they remove selection biases, which are related to stellar luminosities. The level of completeness is limited by a variety of systematic effects and needs to be quantified. Previous efforts to compile volume-complete samples of binaries require to determine the orbital parameters with time-resolved observations in addition and are restricted to very small volumes and stars of certain spectral types or evolutionary stages (e.g. Raghavan et al. 2010, ApJS, 190, 1).

$\rho$ varies drastically dependent on the spectral types and the evolutionary states of the stars. A crucial number to explore evolutionary connections between stars is the occurence rate $o=\rho/\tau$,  where $\tau$ is the lifetime of the star at the given evolutionary stage. $o$ is defined as the number of stars per volume and time, which are in the stage characterised by the observed parameters. For binaries to qualify as progenitors $o_{\rm prog}\geq o_{\rm post-MT}$. If $o_{\rm prog}$ is smaller, those binaries cannot be the exclusive progenitors. Instead there must be other progenitor types contributing to the post-MT population. If $\sum_{i=1}^{n}\,o_{\rm prog, i}$ of all the $n$ predicted progenitor types is still smaller than $o_{\rm post-MT}$, there must a yet unknown channel contributing. Bottom line here is that no matter how complicated a binary evolution network is $\sum_{i=1}^{n}\,o_{\rm prog, i}=\sum_{j=1}^{m}\,o_{\rm post-MT, j}$ in all interaction stages. Or in other words, occurence rates are conserved in binary evolution and are a powerful diagnostics to reveal shortcomings in our understanding of binary evolution such as evolutionary channels not yet accounted for or to identify potential evolutionary origins of exotic and rare types of stars. 

But how can we be sure that a specific binary system will really turn into another one given all the uncertain physics involved? In general is not possible to clearly associate progenitor and post-MT systems. This limits the diagnostic potential of observational studies considerably. {\it Some interactions have a much higher diagnostic potential and allow us to learn much more. Those are like well controlled experiments the Universe conducts for us.} 

In a fully characterised volume-complete sample of progenitors and associated post-MT systems we can determine $o_{\rm prog}$ and $o_{\rm post-MT}$ based on measured $\rho$ and theoretical $\tau$. Since we also know the $q$ values of the progenitors, we can assume that the binaries with the smallest $q$ values will do RLOF. Therefore, $o_{\rm prog}$ should correspond to the occurence rate of the post-RLOF systems $o_{\rm pRLOF}$ for $q\leq q_{\rm crit}$. Or in other words, $q_{\rm crit}$ can be directly constrained by sorting the progenitors with ascending $q$ and assuming that the fraction corresponding to $o_{\rm pRLOF}$ is associated with the small-$q$ tail of the $o_{\rm prog}$-distribution. The highest $q$ reached when filling the distribution is then equal to $q_{\rm crit}$.

{\it In conclusion, volume-complete samples of well characterised progenitor and post-MT systems have a diagnostic potential not only to study the interactions themselves, but also to reveal inconsistencies in the binary evolution tree, and help to understand the complicated connections between the diverse channels. They are needed to get a truly comprehensive picture of stellar evolution.}

\section{Experiments in binary evolution}

The main and on first sight insurmountable caveat with volume-complete samples is that it is impossible to compile them even in the close vicinity of the Sun. To fully characterise stars, spectra of decent quality are needed to determine the atmospheric parameters and derive the fundamental parameters. For binary systems, the orbital parameters need to be measured as well, which requires time-resolved spectroscopy or astrometry. The main limiting factor to compile such samples is the intrinsic faintness of the coolest objects. However, since we are only interested in stars that have interacted within the age of the Universe or less, we can restrict our sample to stars with lifetime of less than a Hubble time and therefore main-sequence (MS) masses $M>0.8\,M_{\rm \odot}$. This already removes $80\%$ of the close-by sources (Kirkpatrick et al. 2024, ApJS, 271, 55). 

These numbers are massively reduced further, if we only focus on binaries, which will interact. For each of the specific progenitor populations studied there is a $M_{\rm donor}-P$ relation, where $M_{\rm donor}$ is the mass of the donor star initiating the mass transfer and $P$ is the orbital period. It is also not necessary to probe the same volumes for different types of stars. Due to the initial mass function cool low-mass MS stars are way more numerous than hot intermediate mass stars and since they live much longer, we see many more of them in this stage. To determine space densities with similar statistical uncertainties, the volumes to be probed for stars with different masses can therefore be quite different.  

Most interactions are not suited to perform the detailed studies described above. Stable mass transfer of MS  stars for example can lead to diverse outcomes, which are highly dependent on the binary parameters and uncertain physics mostly related to stellar evolution and loss of mass and angular momentum (e.g. Nelson \& Eggleton 2001, ApJ, 552, 664). It is therefore not possible to clearly link or even identify progenitor and post-MT systems. The exact properties of the progenitors for MT on the asymptotic giant branch (AGB) are also not well enough constrained for a dedicated survey project. AGB stars exhibit rapid changes in radius during thermal pulses accompanied by dynamical pulsations (Mathieu \& Pols 2025). Due to the more moderate mass loss on the first red giant branch (RGB), the MS progenitors for mass transfer on the RGB can be characterised better. To remove degeneracies related to the donor masses, which show a wide variety in young stellar populations, one can restrict studies to the old stars in the solar neighbourhood mostly associated with the Galactic thick disk and the halo. The turn-off mass in those populations is about $0.8\,M_{\rm \odot}$, which is also the mass of the donors in progenitors of MT. The volume around us with accurate Gaia parallaxes contains hundreds of thousands relatively bright, but old MS stars.

In the old Galactic population we can study the mergers of MS stars by comparing progenitor candidates ($P\sim0.4\,{\rm d}$) with the population of merged blue stragglers. The envelope stripping of such old stars, which then form low-mass (pre-)helium white dwarf (WDs, $M_{\rm He-core}\sim0.2\,M_{\rm \odot}$) or core helium burning sdO/B stars ($M_{\rm He-core}\sim0.47\,M_{\rm \odot}$) requires progenitor binaries with periods of $\sim4\,{\rm d}$ and $\sim500\,{\rm d}$, respectively. For intermediate mass progenitors of sdO/Bs, the He-core mass is even directly correlated with $M_{\rm donor}$ (e.g. Arancibia-Rojas et al. 2024, MNRAS, 527, 11184) and the orbital periods of the progenitors ($\sim30-100\,{\rm d}$) only a function of $M_{\rm comp}$. Also the post-MT systems have characteristic properties, which allow us to identify and study them. Depending on the details of mass transfer, they can have orbital periods ranging from minutes to several years.

{\it Those experiments in binary evolution can be used to put observational constraints on yet unconstrained parameters such as the stability criteria, the efficiencies of the transfers of mass and angular momentum, and the CE efficiency, which are needed to improve our understanding of stellar interactions in general.} To to do this, we need to determine the atmospheric, fundamental and in particular orbital parameters for thousands of binary systems with periods spanning several orders of magnitude.

\section{Technology developments and data processing}

Current and planned massive surveys such as Gaia or 4MOST will determine the atmospheric and fundamental parameters for millions of stars around us. The bottleneck are the orbital parameters, because accurate determinations require a good coverage of the orbital phases and a high radial-velocity (RV) accuracy to measure the small RV variations for the long-period systems. Gaia DR4+ will for sure be a game changer, because it is expected to determine astrometric and spectroscopic solutions for millions of binaries. However, the old MS stars we need to study are much rarer and the respective fractions of progenitor binaries for the interactions of interest at the percent level only (Moe \& di Stefano 2017, ApJS, 230, 15; Moe et al. 2019, ApJ, 875, 61) meaning that Gaia might only find hundreds of them in total. 

Since the patterns of Gaia visits of individual objects are not optimized for binary studies, they imprint a very complex selection function which is needed to estimate the completeness (e.g. El Badry et al. 2024, OJAp, 7, 100). Hot intermediate mass stars with relatively short orbital periods as well as pre-He-WD and sdO/B stars will not be solved by Gaia at all. Also running and upcoming ground-based surveys, which obtain time-resolved spectroscopy (e.g. LAMOST), only take a moderate number of multiple observations at random times, which in many instances only helps to detect variations, but not to solve the orbits.

The next generation of time-resolved spectroscopic surveys should be optimized to follow-up and solve whole populations of binary systems in an efficient way. A key ingredient would be a scheduler, which predicts the best times of the next observation for any given target based not only on the data already taken for this target (e.g. to distinguish between alias periods identified in Fourier space). The scheduler should make use of priors characterising the whole population of those binaries and modify them on the spot as new data is taken. This requires the full processing of the data on the fly and in this way the follow-up can be optimized to check more likely solutions as a priority. Given that the number of binaries of any given population visible on the sky at any given time is rather small, the scheduler needs to do that for many different populations at the same time.

The survey instrument(s) should be capable of taking data at individual epochs and with individual exposure times for each target. Due to the short-period variability of some post-MT systems (hours or even minutes), the exact epochs need to be adjustable dynamically in real time. The field of view needs to be as large as possible to maximize the number of targets. Due to the brightness contrast between the progenitors and post-MT systems as well as the large diversity in periods and RV amplitudes there should be the option to switch between medium- and high-resolution spectroscopy for any given target. Since we aim at probing the wider solar neighbourhood, most targets are bright enough to be observed with 4m-class telescopes. The total duration of the survey should be many years to cover also the longest period systems. A network of independent telescopes would provide most flexibility and the widest field-of-view at any given time. A less efficient alternative might be a multi-object spectrograph with large field-of-view where all the fibres can be exposed and read-out independently.

\end{document}